# Radial trends in Galactic globular clusters and their possible origin

V. Kravtsov[1,2]

[1] SAI, Lomonosov Moscow State University, Moscow, Russia
[2] INCT, Universidad de Atacama, Copoiapo, Chile

Contact / vkravtsov1958@gmail.com

**Resumen** / El tiempo de relajacion al radio de semiluz de los cumulos globulares (CGs) de la Galaxia es normalmente dentro de varios Ga ($10^9$ años). Entonces se espera que la myoria de CGs sean bastante bien relajados, tomando en cuenta su edad alrededor de 12-13 Ga. O sea cualquier segregacion radial inicial entre estrellas de la misma masa inicial en la secuencia principal (SP), particularmente entre los progenitores de las actuales estrellas pertenecientes a las ramas de sub-gigantes y gigantes rojas (RSG, RGR), ya deberia desapareser. Sin embargo, se han acumulado pruebas que contradicen a esta expectativa. La paradoja se podria resolver tomando en cuenta el efecto de colisiones estelares que ocurren en los nucleos densos de CGG. La tasa de colisiones es particularmente elevada mientras los nucleos estan colapsando. Esta formacion de nuevas estrellas hace a las partes centrales ser no relajadas. Se llama la atencion que los rezagados azules de colision observados actualmente en CGs serian de muchpo menor numero que el de sus homologos de menor masa acumulados en la SP durante el tiempo de la vida de CGs. El efecto de este proceso seria que las estrellas de la misma luminosidad en las SP/RSG/RGB no seran de la misma masa sino de un rango no despreciable.

**Abstract** / The relaxation time at the half-mass radius of Galactic globular clusters (GGCs) is typically within a few Gyr. Hence, the majority of GGCs are expected to be well relaxed systems, given their age is around 12-13 Gyr. So any initial radial segregation between stars of the same initial mass on the main sequence (MS), in particular, the progenitors of the present day sub-giant and red-giant branch (SGB, RGB) stars should already have dissipated. However, a body of evidence contradicting to these expectations has been accumulated to date. The paradox could be solved by taking into account the effect of stellar collisions. They occur at particularly hight rate in collapsing nuclei of GGCs and seem to be mainly responsible for unrelaxed central regions and the radial segregation observed. We draw attention that actually observed collisional blue stragglers should be less numerous than their lower-mass counterparts formed and accumulated at and below the present day MS turnoff. The effect of this is that MS/SGB/RGB stars of a given luminosity are not of the same mass but fall in a range of mass.

Keywords / globular clusters: general --- Hertzsprung-Russell and C-M diagrams --- stars: low-mass

## 1. Introduction

There is strong evidence that stellar populations in Galactic globular clusters (GGCs) exhibit dissimilari-ties in comparison with stellar populations in Galac-tic open clusters and in the field. The key one is that anti-correlations between abundances of some chemical elements, in particular, O-Na anti-correlation, are not observed among field and Galactic open cluster stars. The apparent manifestations of multiple stellar popula-tions are not obviously observed (at least in statistical sense) in open clusters, too. Therefore, we should search for those factors which could be responsible for (or the main contributors to) these dissimilarities. One of the most probable factors is assumed to be stellar collisions which do occur in the densest part of GGCs but are of much lower probability or improbable at all in other environments, i.e. in open clusters and in the field.

## 2. Radial segregation in globular clusters

One of the intriguing and controversial issues of multiple stellar populations in GGCs is whether there is any signi cant difference in radial distribution of the sub-populations in their parent globular clusters. Probably, Norris Freeman (1979) were rst who reported on difference in the radial distribution of red giant branch (RGB) stars with different CN strength in a GGC, namely in 47 Tuc. Radial color variations were later recovered in GGCs in the sense that the clusters were found to be bluer, in addition to the strengthening of Balmer ab-sorption lines, towards the cluster centers. These ef-fects were summarized and discussed by Djorgovski & Piotto (1993). It has been found that most of these GCs are post-core-collapse objects and the gradient typically extends to at least a few tenths of arcsec in radius and even up to 100" arcsec or further. Carretta et al. (2009) statistically achieved generalized conclusion about different spatial distributions of the spectroscopi-cally distinguished components of stellar populations in a sample of 15 GGCs. Radial segregation between the sub-samples of both SGB and RGG stars with different UV-based photometric characteristics was found in a number of GGCs (Kravtsov et al., 2010a,b, 2011; Lardo et al., 2011). HST photometry in the central parts of two metal-rich GGCs, NGC 6388 and NGC 6441, revealed (Bellini et al., 2013) a probable difference in the radial distribution of two distinct stellar sub-populations in each of the clusters. Larsen et al. (2015) found a radial trend of the ratio of primordial to nitrogen-enhanced giants in the GGC M15, which changes its sign at the radial distance near the half-light radius. Similarly, Lim et al. (2016) found more centrally concentrated CN-weak red giants compared to the CN-strong ones, in the GGCs NGC 363 and NGC 6723. Recently, Dalessandro et al. (2018) reported that first-generation stars in the GC M80 are significantly more centrally concentrated than extreme second-generation stars out to 2.5 half-light radius from the cluster center. It is important that these radial effects were revealed not only among SGB and RGB stars, but also among MS turnoff (MSTO) stars in 47 Tuc (Kucinskas et al., 2014).

An important point here is that the proposed scenarios of the origin of sub-populations in GGCs proceed from observational fact that the sub-populations are old and nearly coeval within a few $10^8$ yr. This implies very unlikely radial segregation between the sub-populations in the majority of GGCs for the following reason. Galactic GGCs are about 12-13 Gyr old while their relaxation time at the half-mass radius is typically within a few Gyr (Harris, 1996). Therefore, the majority of GGCs have to be well relaxed systems. This implies that any initial or very old radial segregation between stars of the same or slightly different mass on the main sequence (MS) should already have dissipated. A portion of these primordial MS stars were, in particular, the progenitors of the present day SGB and RGB stars. This means that stars actually located on these branches at the same luminosity level are expected to exhibit no radial segregation between them. However, a body of evidence contradicting to the latter expectation has been accumulated to date and demonstrated above.

One can assume that the observed radial effects are due to fairly different mass among SGB/RGB/MSTO stars or/and unrelaxed effects caused by recently occurred processes in the innermost parts of GGCs. We (Kravtsov, 2017) recently already suggested that a radial trend of the mass of RGB stars might mimic the iron abundance trend in NGC 3201 we found earlier (Kravtsov, 2013) in this and some other GGCs. Here in developing these assumptions we consider the process that could mainly be responsible for the radial trends/segregation.

## 3. Lower-mass counterparts of collisional blue stragglers

Collision of MS stars is thought to be the most efficient mechanism, alternative to mass transfer in binary systems, of blue straggler (BS) formation in the densest parts of GGCs. Collisional BSs (CBSs) were really revealed in GGCs (Ferraro et al., 2009; Dalessandro et al., 2013). However, its role cannot be reduced to this particular outcome. It should be considered in more general context as a mechanism inducing active star formation in GGCs and leading to the formation and accumulation, during 12-13 Gyr, of populations of newly formed MS stars, lower-mass counterparts of BSs. They should be much more numerous than the observed CBSs because of longer lifetime and higher rate of their formation as compared to CBSs.

Mass transfer in binary systems, where it is allowed between MS stars among their various possible combinations, forms new modified MS stars, the majority of which are mass-transfer BS (MTBSs) located above the MSTO (Stepien & Kiraga, 2015). Their lower-mass counterparts formed simultaneously with MTBSs are less numerous and fall in narrower mass range than lower-mass counterparts of CBSs, but they are added to the population of latter MS stars in the upper MS and thereby increase the total population of modified MS stars below the MSTO and the proportion modified-to-primordial MS stars. This proportion is presumably increased with increasing GC age. We note here that the main "reservoir" of MS stars, potential targets of collision, is at $M < 0.45$ Msol. The portion of MS stars in this mass range depends on the initial mass function of GGCs, but the number of primordial MS stars in this mass range is roughly ten times the number of MS stars with mass $0.45$ Msol $< M < 0.85$ Msol. Therefore, the rate of stellar collisions in the mass range $M < 0.45$ Msol with formation of lower-mass counterparts of BSs can be many times as high as that in the mass range $0.45$ Msol $< M < 0.85$ Msol (Davies, 2016) that results in the formation of CBSs.

Some important sequences of the formation and accumulation of lower-mass counterparts of BSs are as follows.

First, it obvious that the MS, in general, and its upper part, in particular, should consist of stars of different mass at a given luminosity but not of the same mass. The range of mass near the present day MSTO may be as large as $M \approx 0.10$ Msol. Stellar models show that, for example, a star of the mass 0.8 Msol and [Fe/H]= -1.3, formed 2 Gyr ago from two collided primordial MS stars of the mass 0.4 Msol each one, would be now fainter, by nearly 1.0 mag. in the V band, than its primordial counterpart of the same mass, which is approaching to the present day MSTO.

Second, very important point is that the new MS stars formed by collision are expected to have modified superficial CNO abundance as compared with that of primordial MS stars (Glebbeek et al., 2008).

Third, primordial stars of mass M and lower-mass counterparts of BSs of the mass(es) M + ΔM (because of higher hydrogen abundance in their nuclei) will evolve from the MS to a given luminosity on the RGB on the same timescale. That is the RGB should also harbor stars of a range of mass at a given luminosity, ΔM being larger on the RGB than on the MS due to contribution of evolved BSs themselves.